\documentclass[
nofootinbib,
noeprint,
amsmath,amssymb,
aps,
nobibnotes,
twocolumn,
floatfix,
]{revtex4-2}

\usepackage[utf8]{inputenc}
\usepackage{graphicx}
\usepackage{amsmath}
\usepackage{amssymb}
\usepackage{bm}
\usepackage{txfonts}
\usepackage{multirow}
\usepackage{mathptmx} 
\usepackage{siunitx}
\usepackage{xargs}
\setcitestyle{super}

\usepackage[pdftex,dvipsnames]{xcolor}
\usepackage[colorlinks=true, urlcolor=blue, linkcolor=blue, citecolor=blue]{hyperref}
\usepackage{xcolor}
\usepackage{braket}
\usepackage{soul}
\usepackage{verbatim}

\begin{document}
\title{Squeezing, trisqueezing, and quadsqueezing in a spin-oscillator system}
\date{\today}
\author{O. B\u{a}z\u{a}van, S. Saner, D. J. Webb, E. M. Ainley, P. Drmota, D. P. Nadlinger, G. Araneda, D. M. Lucas, C. J. Ballance, R. Srinivas\\
\normalsize{Department of Physics, University of Oxford, Clarendon Laboratory, Parks Road, Oxford OX1 3PU, United Kingdom \\Email: oana.bazavan@physics.ox.ac.uk}
}


\begin{abstract}
Quantum harmonic oscillators model a wide variety of phenomena ranging from electromagnetic fields to vibrations of atoms in molecules. Their excitations can be represented by bosons such as photons, single particles of light, or phonons, the quanta of vibrational energy. Linear interactions that only create and annihilate single bosons can generate coherent states of light~\cite{glauber1963coherent} as well as coherent states of motion of a confined particle~\cite{wineland1998experimental}. Introducing $n^{\rm th}$-order nonlinear interactions, that instead involve $n$ bosons, enables increasingly complex quantum behaviour. For example, second-order interactions enable squeezing~\cite{walls1983squeezed}, which can be used to enhance the precision of measurements beyond classical limits~\cite{caves1981quantum}, while third (and higher)-order interactions create non-Gaussian states which are essential for continuous-variable quantum computation~\cite{lloyd1999quantum}. However, generating nonlinear interactions is challenging as it typically requires either higher-order derivatives of the driving field~\cite{bloembergen1982nonlinear}, which become exponentially weak, or specifically designed hardware\cite{mielenz2016arrays, hillmann2020universal, frattini2017wave}. Hybrid systems, where an oscillator is coupled to an additional spin degree of freedom, provide a solution. In these systems, ranging from trapped ions~\cite{monroe1996schrodinger}, atoms~\cite{haroche2013nobel}, superconducting qubits~\cite{blais2004cavity}, to diamond colour centres~\cite{evans2018photon}, linear interactions that couple the oscillator to the spin are readily available. Here, using the internal spin states of a single trapped ion coupled to a single bosonic mode of its motion, we use two of these linear interactions mediated by their spin to demonstrate a new class of nonlinear bosonic interactions~\cite{sutherland2021universal} up to fourth-order. In particular, we focus on generalised squeezing interactions~\cite{braunstein1987generalised} and demonstrate squeezing, trisqueezing, and quadsqueezing. We characterise these interactions, including their spin dependence and unitarity, and perform full-state tomography by reconstructing the Wigner function of the resulting states. We also discuss the scaling of the interaction strength, where we are able to drive the quadsqueezing interaction more than 100 times faster than using conventional techniques. Our method presents no fundamental limit in the interaction order $n$ and applies to any platform that supports spin-dependent linear interactions. Strong higher-order nonlinear interactions unlock the study of fundamental quantum optics, quantum simulation, and computation in a hitherto unexplored regime.

\end{abstract}
\maketitle

\begin{figure}[ht!]
    \centering
    \includegraphics[width=\linewidth]{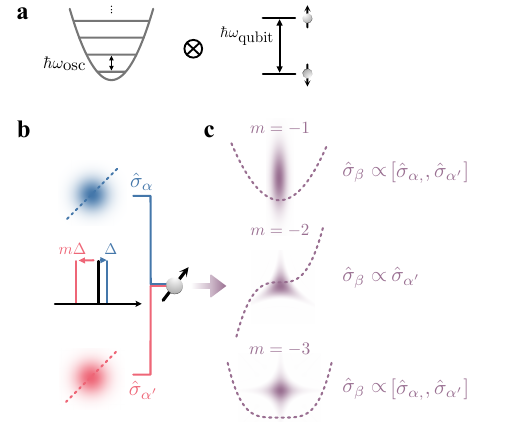}
    \caption{Conceptual illustration of spin-mediated nonlinear interactions. \textbf{a}, Hybrid spin-oscillator system. The protocol requires a quantum harmonic oscillator with energy splitting $\hbar\omega_\textrm{osc}$ (left) coupled to a spin system with energy splitting $\hbar\omega_\textrm{qubit}$ (right). \textbf{b}, Frequency settings for spin-dependent linear interactions. We apply two spin-dependent forces which are detuned from the oscillator motion frequency $\omega_\textrm{osc}$ by $\Delta$ and $m\Delta$, where $m$ is an integer. These interactions are linear and cause a spin-dependent displacement. We set the spin components of these forces $\hat{\sigma}_\alpha$ and $\hat{\sigma}_{\alpha'}$ such that they do not commute, i.e. $[\hat{\sigma}_\alpha, \hat{\sigma}_\alpha'] \neq 0$. We show the Wigner functions of the coherent states (blue and red blobs) that would be generated by the effective potential of the linear interactions (blue and red dashed lines). \textbf{c}, Generation of nonlinear interactions. By adjusting the relative detunings of the linear interactions, and hence $m$, we can drive arbitrary nonlinear interactions. Setting $m=-1$ gives rise to squeezing $\sim  (\hat{a}^{\dagger 2}+\hat{a}^2)$, $m=-2$ trisqueezing $\sim  (\hat{a}^{\dagger 3}+\hat{a}^3)$, and $m=-3$ quadsqueezing $\sim (\hat{a}^{\dagger 4}+\hat{a}^4)$. Purple dashed lines indicate the effective potential for the nonlinear interactions that are proportional to $(\hat{a}^{\dagger}+\hat{a})^n$; by setting $m = 1-n$ we can select the terms in the expansion of this potential that correspond to generalized squeezing interactions. The Wigner functions of the corresponding generalized squeezed states are overlayed on top in purple.}
    \label{fig:theory-fig1}
\end{figure}

\begin{figure*}[ht!]
    \centering
    \includegraphics[width=\linewidth]{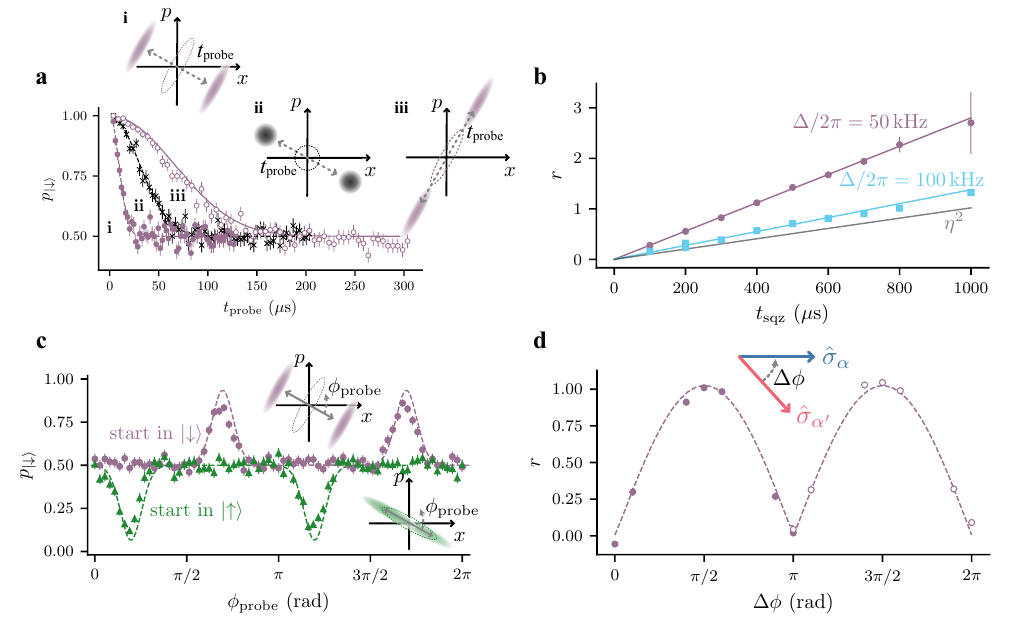}
    \caption{Characterisation of the spin-dependent squeezing interaction. After applying the squeezing interaction, we employ a probe spin-dependent force (SDF) to map the oscillator state to the spin state ($p_{\ket{\downarrow}}$). The insets show the effect of the probe SDF on the Wigner functions of the respective motional states in phase space. The dashed ellipses represent the state before the probe SDF is applied. \textbf{a}, Inferring the squeezing parameter $r$. As we vary the probe SDF duration, $t_\textrm{probe}$, the spin-dependent displacement separates the oscillator wavefunction (insets). The probe SDF is applied along the two principal axes of a squeezed state (\textbf{i, iii}) and to a thermal state close to the motional ground state (\textbf{ii}). We deduce $r = 1.09(4)$ by fitting (dashed lines) the data in \textbf{i} and \textbf{ii}. For splitting about the anti-squeezed axis (\textbf{iii}), we plot the numerical simulation (continuous line) which accounts for motional decoherence effects. \textbf{b}, Detuning dependence of interaction strength. We plot $r$ against $t_{\rm sqz}$ for $\Delta/2\pi=\SI{50}{\kilo \hertz}$ and $\Delta/2\pi=\SI{100}{\kilo \hertz}$. The theory lines (continuous purple and cyan lines) are given by $r=\Omega_2t_{\rm sqz}$. The grey continuous line illustrates the squeezing parameter that would be achieved by driving the second-order spatial derivative of the field with the same laser power. \textbf{c}, Spin dependence of interaction. We apply the probe SDF for a fixed duration and vary its phase $\phi_{\rm probe}$. The dashed lines are fits to the data. Peaks and dips occur when the probe SDF is aligned with the anti-squeezed axis. Changing the initial spin state from $\ket{\downarrow}$ to $\ket{\uparrow}$ shifts the pattern by $\pi/2$, confirming the spin dependence of the interaction.~\textbf{d}, Non-commutativity of the interaction SDFs. We apply the two interaction SDFs with bases $\hat{\sigma}_\alpha = \hat{\sigma}_{\phi}$ and $\hat{\sigma}_{\alpha’} = \hat{\sigma}_{\phi+\Delta\phi}$ and measure $r$ as a function of $\Delta\phi$. When the forces commute ($\Delta\phi = 0,\ \pi,\ {\rm and}\ 2\pi$) the initial state is not squeezed $r = 0$ and when the forces do not commute ($\Delta\phi = \pi/2\ {\rm and}\ 3\pi/2$), the squeezing is maximised.
    The data is fit using $A|\sin{{\Delta\phi}}|$, where $A$ is a constant (dashed lines). The change in marker (full to empty circles) reflects the change in the phase of the probe SDF (see main text). Error bars on the data points indicate 68\% confidence intervals from the shot noise for ({\bf a}, {\bf c}) or based on the fit employed to estimate $r$ ({\bf b}, {\bf d}). The error bars are occasionally smaller than the marker size.
}
    \label{fig:sqz_charact-fig2}
\end{figure*}

Nonlinear processes and interactions in quantum harmonic oscillators are ubiquitous in various technological and scientific applications, ranging from frequency conversion~\cite{fejer1994nonlinear} and nonlinear spectroscopy~\cite{bloembergen1982nonlinear} to the creation of nonclassical states like entangled photon pairs~\cite{kwiat1995new} and squeezed states~\cite{walls1983squeezed}. Squeezed states, which are generated by second-order bosonic processes, reduce the uncertainty in one observable, such as position, while increasing it in its conjugate, momentum~\cite{caves1981quantum}. Such states have been used for enhancing the sensitivity of gravitational wave detectors~\cite{aasi2013enhanced}, microscopy~\cite{casacio2021quantum}, and the measurement of small electric fields~\cite{burd2019quantum}. Beyond conventional squeezing, there has been a long-standing interest in higher-order generalised squeezing interactions~\cite{braunstein1987generalised} as they exhibit non-Gaussian statistics and nonclassical properties such as Wigner negativity~\cite{banaszek1997quantum, albarelli2018resource, takagi2018convex}. For example, trisqueezing (a third-order interaction) is a non-Gaussian operation which is a required resource for continuous variable quantum computation (CVQC)~\cite{zheng2021gaussian}. Together with Gaussian operations, such as displacement and squeezing, it enables computational universality and error correction~\cite{lloyd1999quantum, gottesman2001encoding, braunstein2005quantum, de2022error, sivak2023real}.

However, realising these nonlinear interactions faster than decoherence mechanisms has posed experimental challenges, especially as the interaction strength diminishes with increasing order. Generating any one of these interactions typically requires careful hardware considerations such as specifically tailored ion trap geometries~\cite{mielenz2016arrays} or the design of superconducting microwave circuits~\cite{hillmann2020universal, frattini2017wave}. 
For example, while squeezing of a harmonic oscillator has been demonstrated using electromagnetic fields~\cite{slusher1985observation}, mechanical oscillators~\cite{wollman2015quantum}, and trapped ions~\cite{meekhof1996generation}, trisqueezing has only recently been demonstrated by Refs.~\citenum{chang2020observation, eriksson2023universal} in superconducting microwave circuits. Engineering higher than third-order bosonic interactions has thus far been an outstanding challenge.

Instead of creating direct bosonic interactions, hybrid spin-boson systems offer an additional degree of freedom which can be used to mediate effective interactions. In such systems, the oscillator can be coupled to the spin via a spin-dependent interaction that is linear in the bosonic mode. These interactions are readily available in a variety of platforms and used extensively to realise boson-mediated spin-spin entanglement that overcomes the intrinsically weak spin-spin interactions~\cite{cirac1995quantum, sorensen1999quantum, sorensen2000entanglement}. Here, following the proposal in Ref.~\citenum{sutherland2021universal}, we instead use the spin to mediate bosonic interactions. Focusing on generalised squeezing, we drive two of these linear spin-dependent interactions concurrently to demonstrate up to fourth-order bosonic interactions using a single trapped ion whose motion is a harmonic oscillator that can be coupled to its internal spin states. Notably, we use the same two linear interactions to create squeezing, trisqueezing, and quadsqueezing by simply adjusting the interaction frequency. \\

To elucidate how we generate these $n^{\rm th}$-order interactions, we first consider the quantum harmonic oscillator, which can be described by the operators $\hat{a}^\dagger$ and $\hat{a}$ that create and annihilate a boson, respectively. In hybrid systems, as shown in  Fig.~\ref{fig:theory-fig1}\textbf{a}, this oscillator can be coupled to a spin via a spin-dependent force (SDF) described by the interaction Hamiltonian

\begin{equation}
    \begin{split}
    \hat{H}_\textrm{SDF} =&\ \frac{\hbar \Omega_{\alpha}}{2} \hat{\sigma}_{\alpha} (\hat{a}e^{-i(\Delta\, t+\phi_\alpha)} + \hat{a}^\dagger e^{i(\Delta\, t+\phi_\alpha)}), \\
    \end{split}
\label{eq:1sdf}
\end{equation}

which is linear in $\hat{a}^\dagger$ and $\hat{a}$.
This type of interaction can be generated in many systems such as photons in a microwave cavity coupled to a superconducting qubit~\cite{blais2004cavity}, or phonons coupled to the internal spin state of trapped ions~\cite{monroe1996schrodinger}. The coupling to the spin is described by the Hermitian operator $\hat{\sigma}_{\alpha}$, which is a linear combination of the Pauli operators $\hat{\sigma}_{x, y, z}$. The SDF results in a displacement of the harmonic oscillator state, conditioned on the spin state. This displacement depends on the interaction strength $\Omega_\alpha$, as well as $\Delta$ and $\phi_\alpha$, which are the detuning and phase, respectively, of the SDF relative to the harmonic oscillator with frequency $\omega_\textrm{osc}$.

The nonlinear spin-dependent interactions we seek to generate are the generalised squeezing interactions~\cite{braunstein1987generalised} described by
\begin{equation}
    \hat{H}_\textrm{NL} = \frac{\hbar\Omega_n}{2}\hat{\sigma}_{\beta}(\hat{a}^ne^{-i\theta}+\hat{a}^{\dagger n}e^{i\theta}),  
\label{eq:generalised_sqz}
\end{equation}
where $n$ is the order of the interaction, $\Omega_n$ its strength, and $\theta$ the axis of the interaction. 
For $n=2, 3, 4$, this corresponds to spin-dependent squeezing, trisqueezing, and quadsqueezing, respectively. Applying the Hamiltonian in Eq.~\eqref{eq:generalised_sqz} for a duration $t_{\rm sqz}$ generates $n^{ \rm th}$-order squeezed states characterised by the squeezing parameter ${r = \Omega_n t_{\rm sqz}}$.
Using conventional techniques, the higher the order of the interaction, the more demanding it is to generate. For example, in trapped ions, these interactions are conventionally driven by higher-order spatial derivatives of the electro-magnetic field~\cite{meekhof1996generation}, where $\Omega_n$ varies with $\eta^n$. The Lamb-Dicke parameter $\eta$ corresponds to the ratio of the ground-state extent of the ion ($\sim$~10\,nm) to the wavelength of the driving field ($\sim$~500\,nm). Thus, $\eta$ is usually small and every subsequent order is weaker by more than an order of magnitude. This unfavorable scaling holds not only for trapped ions but also for other platforms such as superconducting circuits~\cite{mundhada2019experimental}.

Here, we circumvent this scaling by instead combining two non-commuting SDFs, each of which is linear. Together, they generate a plethora of nonlinear interactions with different resonance conditions, as proposed in Ref.~\citenum{sutherland2021universal} (see Fig.~\ref{fig:theory-fig1}\textbf{b}-{\bf c}).
The interaction is then described by 
\begin{equation}
    \begin{split}
    \hat{H} =&\  \frac{\hbar\Omega_{\alpha}}{2} \hat{\sigma}_{\alpha} (\hat{a}e^{-i \Delta\, t} + \hat{a}^\dagger e^{i\Delta\, t}) \\
    &+ \frac{\hbar\Omega_{\alpha'}}{2} \hat{\sigma}_{\alpha'} (\hat{a}e^{-i( m\Delta\, t +\phi_{\alpha'})} + \hat{a}^\dagger e^{i(m\Delta\, t + \phi_{\alpha'})}),
    \end{split}
\label{eq:2sdfs}
\end{equation}
where $\Delta$ and $m\Delta$, with $m$ an integer, are the detunings 
from $\omega_\textrm{osc}$. Without loss of generality, we set $\phi_\alpha = 0$. If the spin components of the two forces do not commute, i.e. $[\hat{\sigma}_{\alpha}, \hat{\sigma}_{\alpha'}]\neq 0$, we can choose $m=1-n$ to satisfy the resonance condition for creating effective interactions corresponding to Eq.~\eqref{eq:generalised_sqz}. For $m=-1, -2, -3$, we generate squeezing, trisqueezing, and quadsqueezing interactions, respectively. The spin dependence $\hat{\sigma}_\beta$ is given by the initial choice of $\hat{\sigma}_{\alpha,\alpha'}$ and the desired squeezing order $n$.
The even-order interactions have a spin dependence which follows as $\hat{\sigma}_\beta \propto [\hat{\sigma}_\alpha, \hat{\sigma}_{\alpha'}]$, while the odd orders follow $\hat{\sigma}_\beta \propto \hat{\sigma}_{\alpha'}$. Hence, by being able to generate SDFs conditioned on any Pauli operator, the spin component of the nonlinear interaction can be arbitrarily chosen. 
The axis $\theta$ of the resulting interaction (Eq.~\eqref{eq:generalised_sqz}) can be controlled by adjusting the SDF phase $\phi_{\alpha'}$. The strength of generalised squeezing $\Omega_n$ is proportional to $\Omega_{\alpha'}\Omega_{\alpha}^{n-1}/\Delta^{n-1}$.
Importantly, and contrary to previous implementations~\cite{meekhof1996generation}, $\Omega_n$ is linear in $\eta$ for all orders $n$ as the detuning $\Delta$ is a free parameter. This scaling applies even though both $\Omega_\alpha, \Omega_{\alpha'}\propto\eta$ because $\Delta$ can be adjusted to keep the overall dependence of $\Omega_n$ linear with $\eta$ (see Supplementary Information).

We experimentally demonstrate these interactions on a trapped $^{88}$Sr$^+$ ion in a 3D radiofrequency Paul trap~\cite{thirumalai2019high}. 
Our qubit states comprise the $\ket{5S_{1/2},\,m_j = -\frac{1}{2}}\equiv\ket{\downarrow}$ and $\ket{4D_{5/2},\,m_j = -\frac{3}{2}}\equiv\ket{\uparrow}$ sublevels of the ion's electronic structure, where $m_j$ is the projection of the total angular momentum along the quantization axis defined by a 146\,G static magnetic field. Aside from the internal (spin) degree of freedom, the ion also vibrates in three dimensions; the harmonic oscillator is defined by the motional mode along the trap axis, with ${\omega_\textrm{osc}/2\pi \approx\SI{1.2}{\mega \hertz}}$ (see Fig.~\ref{fig:theory-fig1}\textbf{a}). We initialise this oscillator close to the ground state with $\bar{n}_{\rm osc}=0.09(1)$.

For creating the nonlinear interactions, we use two SDFs, as described in Eq.~\eqref{eq:2sdfs}, following the M{\o}lmer-S{\o}rensen (MS) type scheme~\cite{sorensen2000entanglement}. Each SDF requires a bichromatic field composed of two tones that are symmetrically detuned from the qubit transition $\omega_\textrm{qubit}$, driven by a \SI{674}{\nano \meter} laser. If the tones are detuned by $\approx \pm\omega_\textrm{osc}$, the spin component of the force is $\hat{\sigma}_{\phi} = \cos\phi \hat{\sigma}_x + \sin\phi \hat{\sigma}_y$, where $\phi$ is given by the mean optical phase of the two tones at the position of the ion. Alternatively, we can obtain a $\hat{\sigma}_z$ spin component by setting the detuning to be $\approx \pm \omega_\textrm{osc}/2$ \cite{roos2008ion, Bazavan2022synthesizing}.
We actively stabilise the optical phase between the laser beams that give rise to the SDFs in order to maintain their non-commuting relationship throughout the experiment.
In our setup, the beam waist radius is 20~µm and the Lamb-Dicke parameter is $\eta=0.049(1)$. 
If the interaction SDF is in the $\hat{\sigma}_\phi$ basis, then its strength is $\Omega_{\alpha, \alpha'}/2\pi\approx \SI{4.6}{\kilo\hertz}$ (laser power \SI{0.5}{\milli\watt}) or $\approx \SI{6.5}{\kilo \hertz}$ (laser power \SI{1}{\milli\watt}). In the $\hat{\sigma}_z$ basis, its strength is $\approx \SI{1.3}{\kilo \hertz}$ (laser power \SI{1}{\milli\watt}).
Moreover, to ensure that the effective Hamiltonian of the resulting nonlinear interactions tends to the ideal Hamiltonian in Eq.~\eqref{eq:generalised_sqz}, we ramp the two bichromatic fields on and off with a sin$^2$ pulse shape. The ramp duration $t_{\rm ramp}$ should be long compared to $1/\Delta$. We characterise the oscillator states generated through the nonlinear interactions by applying a probe SDF on resonance with $\omega_\textrm{osc}$. The probe SDF is also created using an MS scheme. We present complete details of the experimental setup in the Supplementary Information.

We first use this technique to generate spin-dependent squeezing ($n=2$ in Eq.~\eqref{eq:generalised_sqz}) and verify key characteristics of this interaction family: magnitude, spin dependence, and non-commutativity, as shown in Fig.~\ref{fig:sqz_charact-fig2}. These interactions are also unitary, which we investigate in the Supplementary Information.
We set the detunings of the SDFs to be $\Delta$ and $-\Delta$, respectively i.e. $m=-1$. The spin components of the two SDFs are set to $\hat{\sigma}_\alpha = \hat{\sigma}_\phi$ and $\hat{\sigma}_{\alpha'} = \hat{\sigma}_{\phi+\pi/2}$, respectively. The spin basis of the squeezing is thus $[\hat{\sigma}_\alpha, \hat{\sigma}_{\alpha'}] \propto \hat{\sigma}_z$. If we start in $\ket{\downarrow}$ or $\ket{\uparrow}$ (eigenstates of $\hat{\sigma}_z$) the spin component remains unchanged and the squeezing axis depends on the spin state. Once the squeezed state is created, we apply a probe SDF with the spin component in the $\hat{\sigma}_x$ basis with eigenstates $\ket{\pm} = (\ket{\uparrow}\pm \ket{\downarrow})/\sqrt{2}$. Hence, the probe SDF displaces the $\ket{+}$ and $\ket{-}$ components of the resulting state in opposite directions~\cite{lo2015spin}, see insets in Fig.~\ref{fig:sqz_charact-fig2}\textbf{a}. The overlap of the two parts of the harmonic oscillator wavefunction is mapped onto the spin, whose state probability $p_{\ket{\downarrow}}$ is measured by fluorescence readout. We apply the probe SDF for variable durations $t_\textrm{probe}$; as $t_\textrm{probe}$ is increased, the overlap reduces and $p_{\ket{\downarrow}}\rightarrow 0.5$.

As shown in Fig.~\ref{fig:sqz_charact-fig2}\textbf{a}, applying the probe along the squeezing axis (\textbf{i}) reduces the overlap faster than applying the probe orthogonal to the squeezing axis (anti-squeezed axis, \textbf{iii}). We determine the magnitude of the squeezing parameter~\cite{lvovsky2015squeezed} $r$ by fitting the splitting dynamics of a squeezed (\textbf{i}) and the initial thermal state with $\bar{n}_{\rm osc}=0.09(1)$ (\textbf{ii}), where the latter is used to calibrate the magnitude of the probe SDF. The inferred $r= 1.09(4)$, equivalent to \SI{9.5(3)}{\dB} of squeezing.
Extracting $r$ from \textbf{iii} using the analytic model underestimates the value of $r$ due to motional decoherence, whose effect is more apparent in this case as it takes longer to reduce the overlap completely. Nonetheless, the resulting dynamics agree well with numerical simulations that incorporate the motional decoherence. The squeezed state considered here is created by using \SI{0.5}{\milli\watt} for driving each interaction SDF, setting $\Delta/2\pi= \SI{50}{\kilo \hertz}$ and applying the interaction for a pulse duration of $t_\textrm{sqz}= \SI{400}{\micro \second}$ with a ramp duration of $t_\textrm{ramp} = \SI{40}{\micro \second}$~\cite{fnpulseshape}. 

The squeezing parameter of the squeezed state is $r=\Omega_2t_\textrm{sqz}$, where $\Omega_2=\Omega_\alpha\Omega_{\alpha'}/\Delta$ following Eq.~\eqref{eq:generalised_sqz}. We verify this dependence in Fig.~\ref{fig:sqz_charact-fig2}\textbf{b} where we plot $r$ as a function of $t_\textrm{sqz}$ for $\Delta/2\pi = \SI{50}{\kilo \hertz}$ and $\Delta/2\pi = \SI{100}{\kilo \hertz}$. The data agree well with the theory, calculated from independently measured values of $\Omega_\alpha, \Omega_{\alpha'}$, and we observe that the magnitude is inversely proportional to $\Delta$. 
We compare the squeezing strength generated by our method to driving the interaction directly using the second-order spatial derivative of the field\cite{meekhof1996generation}. This interaction strength scales with $\eta^2$ and the values were inferred by considering the same total power of \SI{1}{\milli\watt} for both methods.
This underscores that we can adjust the free parameter $\Delta$ in our method to achieve a higher coupling strength than driving the second-order interaction directly.

We next investigate the spin dependence of the interaction as shown in Fig.~\ref{fig:sqz_charact-fig2}\textbf{c}. 
The spin dependence of our interaction is in contrast to spin-independent squeezing achieved by modulating the confinement of the trapped ions~\cite{heinzen1990quantum, burd2019quantum, wittemer2019phonon}.
We create squeezed states using the same parameters as Fig.~\ref{fig:sqz_charact-fig2}\textbf{a}, and fix the probe SDF duration, $t_\textrm{probe} = \SI{53.6}{\micro \second}$. We scan the phase of the probe SDF $\phi_{\rm probe}$ and measure $p_{\ket{\downarrow}}$. Changing this phase influences the direction about which we split the oscillator wavefunction (see inset). The peaks and the dips of the population correspond to splitting about the anti-squeezed axis and has $\pi$ periodicity. There is a $\pi/2$ shift between the two curves as a result of squeezing about orthogonal axes in phase space introduced by the different spin state settings (see insets). 

To generate this family of interactions, the spin components of the SDFs must be non-commuting. We explore this non-commutativity by varying the phase between the spin components of the two SDFs, i.e., one of the forces is $\hat{\sigma}_\alpha = \hat{\sigma}_\phi$ and the other is $\hat{\sigma}_{\alpha'} = \hat{\sigma}_{\phi+\Delta\phi}$. We measure $r$ as a function of $\Delta\phi$, keeping the phase of the probe SDF constant. The squeezing parameter $r$ varies as $\sin{(\Delta\phi)}$ following the commutator relationship $[\hat{\sigma}_{\phi}, \hat{\sigma}_{\phi+\Delta\phi} ]\propto\sin{(\Delta\phi)\hat{\sigma}_z}$, as shown in Fig.~\ref{fig:sqz_charact-fig2}{\bf d}. If the spin components commute, i.e., $\Delta\phi = 0,\ \pi,\ {\rm and}\ 2\pi$, there is no squeezing, while for $\Delta\phi = \pi/2\ {\rm and}\ 3\pi/2$, the commutator of the spin components, and hence the squeezing, is maximised. When $\sin{(\Delta\phi)}$ becomes negative, i.e., $\Delta\phi > \pi$, the axis of squeezing shifts by $\pi/2$; hence, we change the phase of the probe SDF to $\phi_{\rm probe} +\pi/2$ such that  we always split about the squeezed axis.

\begin{figure*}[ht!]
    \centering
    \includegraphics[width=1.\linewidth]{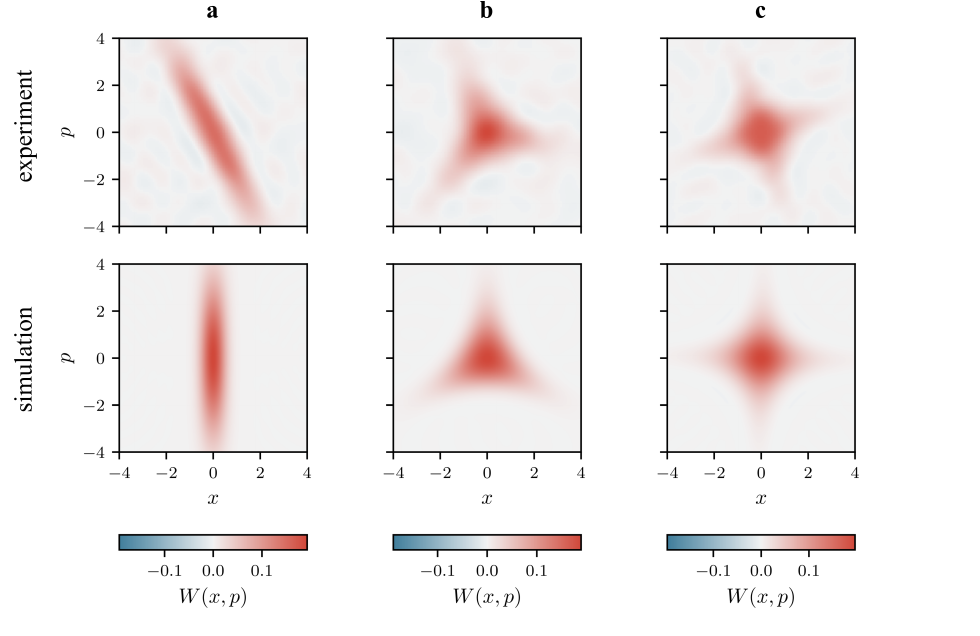}
    \caption{Wigner functions of generalized squeezed states: \textbf{a}, Squeezed state with $r = 1.09(4)$, \textbf{b}, Trisqueezed state with $r_{3s} = 0.19(1)$, \textbf{c}, Quadsqueezed state with $r_{4s} = 0.054(5)$. On the top row, we show Wigner functions $W(x, p)$ reconstructed from experimental data, where $x, p$ are the position and momentum variables associated with the dimensionless position and momentum operators $\hat{x}, \hat{p}$, respectively. The Wigner function is inferred from the measured characteristic function of the oscillator state (see text). On the bottom row, we show Wigner functions of numerically simulated states with independently measured experimental parameters. The rotation observed in comparison to the simulation is due to a constant offset between the squeezing axis $\theta$ and the phase of the probing SDF $\phi_{\rm probe}$. This offset can be calibrated out if desired.}
    \label{fig:wigner}
\end{figure*}

So far, we have focused on squeezed states that have been explored in a variety of platforms. 
Moving to higher-order interactions, we reconstruct the Wigner quasiprobability function~\cite{wigner1932on} of the resulting quantum states to obtain their full description.
 Following Ref.~\citenum{fluhmann2020direct}, we measure the complex valued characteristic function $\chi(\beta) = \braket{\hat{\mathcal{D}}(\beta)}$, where $\hat{\mathcal{D}}(\beta) = e^{\beta\hat{a}^\dagger-\beta^*\hat{a}}$ is the displacement operator and $\beta \in \mathbb{C}$ quantifies the displacement of the oscillator state in phase space. This measurement is an extension of the method discussed in Fig.~\ref{fig:sqz_charact-fig2}, where we apply a probe SDF to split the oscillator wavefunction. Here, we scan both $t_\textrm{probe}$ and $\phi_{\rm probe}$ to obtain the real and imaginary part of the characteristic function, where $\beta \propto t_{\rm probe} \cdot e^{i\phi_{\rm probe}}$ (see Supplementary Information). We then take the 2D discrete Fourier transform of the measured characteristic function $\chi(\beta)$ to obtain the Wigner function $W(x, p)$, where $x, p$ are the position and momentum variables associated with the dimensionless position and momentum operators $\hat{x}, \hat{p}$, respectively. 

We reconstruct the Wigner functions of experimentally implemented squeezed, trisqueezed, and quadsqueezed states and compare them to numerical simulations where the experimental parameters were measured independently. Harnessing the versatility of our method, the trisqueezed and quadsqueezed states were created by simply changing the detuning $m\Delta$. The spin dependence of all the interactions was controlled to be $\hat{\sigma}_z$ and we initialise the spin in the $\ket{\downarrow}$ eigenstate. In Fig.~\ref{fig:wigner}\textbf{a}, we evaluate a squeezed state with $r=1.09(4)$, which is created using the same parameters as in Fig.~\ref{fig:sqz_charact-fig2}\textbf{a}. 

To create the trisqueezed state (Fig.~\ref{fig:wigner}\textbf{b}), we set the detunings of the SDFs to be $\Delta$ and $-2\Delta$, with $\Delta/2\pi = \SI{-25}{\kilo \hertz}$\cite{fnnegdet}. We apply the interaction for $t_{\rm sqz}=\SI{600}{\micro \second}$, with $t_{\rm ramp}=\SI{80}{\micro \second}$. We use a laser power of \SI{1}{\milli\watt} per interaction SDF. We infer the trisqueezing parameter $r_\textrm{3s} = \Omega_3 t_{\rm sqz} = 0.19(1)$ by assuming that the interaction strength follows the theory $\Omega_{\alpha'}\Omega_{\alpha}^{2}/(2\Delta^{2})$ and comparing to simulation (see Supplementary Information).
The basis of the trisqueezing interaction is given by $[\hat{\sigma}_\alpha,[\hat{\sigma}_\alpha, \hat{\sigma}_{\alpha'}]]$. Here, we set the bases of the comprising interaction SDFs to $\hat{\sigma}_\alpha = \hat{\sigma}_\phi$ and $\hat{\sigma}_{\alpha'} = \hat{\sigma}_z$ such that the effective interaction has a $\hat{\sigma}_z$ spin component. 

Lastly, we create quadsqueezed states (Fig.~\ref{fig:wigner}\textbf{c}) by setting the SDF detunings to be $\Delta$ and $-3\Delta$, with $\Delta/2\pi = \SI{25}{\kilo \hertz}$. We apply the interaction for $t_{\rm sqz}=\SI{600}{\micro \second}$, with $t_{\rm ramp}=\SI{80}{\micro \second}$. The laser power used is \SI{1}{\milli\watt} per interaction SDF. Similarly to the trisqueezed state, we determine the quadsqueezing parameter ${r_\textrm{4s} = \Omega_4 t_{\rm sqz} = 0.054(5)}$. The spin basis of the quadsqueezing is given by $[\hat{\sigma}_\alpha,[\hat{\sigma}_\alpha,[\hat{\sigma}_\alpha, \hat{\sigma}_{\alpha'}]]]$. Thus, choosing the basis of the comprising interaction SDFs to be $\hat{\sigma}_\alpha = \hat{\sigma}_\phi$ and ${\hat{\sigma}_{\alpha'} = \hat{\sigma}_{\phi+\pi/2}}$, we again achieve a $\hat{\sigma}_z$ interaction. 

To our knowledge, this is the first implementation of trisqueezing in an atomic system and the first demonstration of a fourth-order bosonic interaction in any platform. Our demonstration has only been possible because the spin-mediated interactions can be significantly enhanced; the quadsqueezing interaction is more than 100 times stronger than an interaction derived from higher-order spatial derivatives of the driving field, assuming the same total laser power (see Supplementary Information). 

Overall, our work explores nonlinear bosonic interactions mediated by the spin, using the same tools that routinely create spin-spin interactions mediated by a bosonic mode in hybrid systems. Using the spin to combine multiple linear interactions, our technique enabled us to demonstrate fourth-order nonlinear interactions without any limit on the achievable order. These interactions would have been otherwise inaccessible using previous techniques. Further, the effective interactions are not limited to only generalised squeezing interactions as shown in this work, but any nonlinear bosonic interaction comprising other combinations of the creation and annihilation operators. Our proof-of-principle demonstration used only a single motional mode of an ion coupled to two of its internal states. Both of these quantum degrees of freedom can be explored further. Firstly, our technique readily extends to multiple modes~\cite{sutherland2021universal} of a single ion or a larger crystal to generate interactions such as the beamsplitter~\cite{brown2011coupled, gan2020hybrid,hou2022coherently}, two-mode squeezing~\cite{metzner2023two}, or cross-Kerr couplings~\cite{ding2017cross}. Such multi-mode interactions are essential for implementing a universal gate set for scalable continuous variable quantum computing\cite{lloyd1999quantum, braunstein2005quantum}. Secondly, the spin dependence of the bosonic interactions creates the enticing possibility of performing mid-circuit measurements on the spin to create resourceful quantum states \cite{kienzler2015quantum, fluhmann2019encoding, drechsler2020state} for quantum simulation, metrology, or error correction. Finally, our technique is also applicable to hybrid spin-boson encodings which are more natural for certain applications such as simulation of quantum field theories~\cite{bermudez2017quantum,bermudez2023synthetic} or nuclear quantum effects~\cite{valahu2023direct} and have the potential of reducing their computational resource requirements~\cite{lee2023fault}.

\section*{Acknowledgements}
We would like to thank A.~Bermudez, M. F. Gely, A. I. Lvovsky, and R. T. Sutherland for insightful discussions.
This work was supported by the US Army Research Office (W911NF-20-1-0038) and the UK EPSRC Hub in Quantum Computing and Simulation (EP/T001062/1). DPN is supported by Merton College, Oxford. GA acknowledges support from Wolfson College, Oxford. OB acknowledges support from Brasenose College, Oxford.  CJB acknowledges support from a UKRI FL Fellowship. RS is funded by the EPSRC Fellowship EP/W028026/1 and Balliol College, Oxford. 

\section*{Author contributions}

OB and SS. led the experiments and analysed the results with assistance from DJW, PD, DPN, GAM and RS; OB, SS, and DJW updated and maintained the experimental apparatus; GAM, EMA rebuilt and maintained the 674 nm laser system with assistance from OB, SS, DJW, PD, and DPN; OB, SS, and RS wrote the manuscript with input from all authors; RS conceived the experiments and supervised the work with support from DML and CJB; DML and CJB secured funding for the work.

\section*{Data Availability}

Source data for all plots are available. All other data or analysis code that support the plots are available from the corresponding authors upon reasonable request.

\section*{Competing Interests}
RS is partially employed by Oxford Ionics Ltd. CJB is a director of Oxford Ionics Ltd. All other authors declare no competing interests.

\bibliographystyle{apsrev4-2}
\bibliography{bibliography}

\begin{thebibliography}{67}
\bibitem [{fnp()}]{fnpulseshape}%
  \BibitemOpen
  \href@noop {} {}\bibinfo {note} {All pulse durations quoted in this text are measured at full-width half-maximum of the pulse shape. The ramp shape is a $\sin(\pi t/2t_{\rm ramp})^2$ with a total rise time given by the ramp duration $t_{\rm ramp}$.}\BibitemShut {Stop}%
\bibitem [{fnn()}]{fnnegdet}%
  \BibitemOpen
  \href@noop {} {}\bibinfo {note} {We choose this negative detuning $\Delta/2\pi = \SI{-25}{\kilo \hertz}$ to avoid off-resonantly driving an interaction corresponding to another motional mode of the ion.}\BibitemShut {Stop}%
\end{thebibliography}
\clearpage
\clearpage
\renewcommand{\thefigure}{S.\arabic{figure}}
\setcounter{figure}{0}
\setcounter{section}{0}
\twocolumngrid
\begin{center}
\vspace{5 mm}
\textbf{\large Supplementary Material}\\
\end{center}

\section{Theory}

\subsection{Spin-mediated nonlinear bosonic interactions} \label{sec:non-int}

In this section, we discuss how we generate effective spin-dependent nonlinear bosonic interactions by applying spin-dependent interactions that are linear in the bosonic creation $\hat{a}^\dagger$ and annihilation $\hat{a}$ operators. The theory was developed in Ref.~\citenum{sutherland2021universal}.

We use spin-dependent forces (SDFs) to generate the required linear interactions. An SDF is described by 
\begin{equation}
    \hat{H}_\textrm{SDF} = \frac{\hbar \Omega_\alpha}{2}\hat{\sigma}_{\alpha} (\hat{a}e^{-i(\Delta t +\phi_\alpha)} + \hat{a}^\dagger e^{i(\Delta t + \phi_\alpha)}), \label{eq_supp:1sdf}
\end{equation}
where $\Delta$ and $\phi_\alpha$ are the detuning and phase of the SDF from the harmonic oscillator with frequency $\omega_{\rm osc}$, respectively. The basis of the spin conditioning $\hat{\sigma}_\alpha$ is a Hermitian linear combination of the Pauli spin-operators $\hat{\sigma}_{x, y, z}$. The strength of the interaction, $\Omega_\alpha$, is proportional to the Lamb-Dicke factor $\eta$. This expression is in the interaction picture with respect to the qubit frequency $\omega_{\rm qubit}$, and the oscillator frequency $\omega_{\rm osc}$, where 
we apply the rotating wave approximation (RWA)
with respect to $\omega_{\rm qubit}$ and $\omega_{\rm osc}$.

As discussed in the main text, the nonlinear interactions are created by applying two SDFs simultaneously, but with different detunings $\Delta$ and $m\Delta$ and bases $\hat{\sigma}_\alpha$ and $\hat{\sigma}_{\alpha'}$. The resulting interaction is 
\begin{equation}
    \begin{split}
    \hat{H} =&\  \frac{\hbar\Omega_{\alpha}}{2} \hat{\sigma}_{\alpha} (\hat{a}e^{-i (\Delta\, t +\phi_{\alpha})} + \hat{a}^\dagger e^{i (\Delta\, t+\phi_{\alpha})}) \\
    &+ \frac{\hbar\Omega_{\alpha'}}{2} \hat{\sigma}_{\alpha'} (\hat{a}e^{-i (m\Delta\, t + \phi_{\alpha'})} + \hat{a}^\dagger e^{i (m\Delta\, t + \phi_{\alpha'})}), 
    \end{split}
    \label{eq_supp:2sdfs}
\end{equation}
where we set $\phi_\alpha = 0$.
To determine the dynamics of the two SDFs, we consider the resulting unitary time propagator found via the Magnus-expansion\cite{magnus1954on, blanes2008the,sutherland2021universal}
\begin{align}
    U(t) &=\mathcal{T}\Big( e^{\frac{-i}{\hbar} \int_0^t \hat{H}(t') dt'}\Big)  \label{eq_supp:magnus}\\
    &\simeq \exp\Big( \frac{-i}{\hbar}\int_0^t dt_1 H_1  \label{eq_supp:1-order}\\
    &- \frac{1}{2\hbar^2}\int_0^t \int_0^{t_1}dt_1 dt_2 [H_1, H_2]  \label{eq_supp:2-order}\\
    &+ \frac{i}{6\hbar^3}\int_0^t \int_0^{t_1} \int_0^{t_2} dt_1 dt_2 dt_3 \left(\left[H_1, \left[H_2, H_3\right]\right] + \left[\left[H_1, H_2\right], H_3\right]\right) \label{eq_supp:3-order}\\
    &+ \frac{1}{12\hbar^4}\int_0^t \int_0^{t_1} \int_0^{t_2}\int_0^{t_3} dt_1 dt_2 dt_3 dt_4 \nonumber\\
    &(
    \left[\left[\left[H_1, H_2\right], H_3\right], H_4\right] + \left[H_1, \left[\left[H_2, H_3\right], H_4\right]\right] \nonumber\\
    &+ \left[H_1, \left[H_2, \left[H_3, H_4\right]\right] \right]
    +\left[H_2,\left[ H_3, \left[H_4, H_1\right] \right] \right])\Big) \label{eq_supp:4-order},
\end{align}
where $\mathcal{T}$ denotes the time-ordering operator and $\hat{H}_k \equiv \hat{H}(t_k)$ is the Hamiltonian describing the system at time $t_k$. The expansion is truncated at the $4^{\rm th}$-order as opposed to $3^{\rm rd}$-order in Ref.~\citenum{sutherland2021universal}.

The first-order term in the Magnus expansion (Eq.~\eqref{eq_supp:1-order}) leads to periodic displacements (i.e.~loops) of the oscillator in phase space. For durations that are integer multiples of $2\pi/\Delta$, the oscillator state returns to its original position. The second term gives rise to a geometric phase, which underlies the effective spin-spin interactions\cite{cirac1995quantum, sorensen1999quantum, sorensen2000entanglement} in trapped ions where the motion mediates the interaction. If $\hat{\sigma}_\alpha, \hat{\sigma}_{\alpha'}$ commute, i.e. ${[\hat{\sigma}_\alpha, \hat{\sigma}_{\alpha'}] = 0}$, the second-order term (Eq.~\eqref{eq_supp:2-order}) is only a scalar corresponding to this geometric phase, causing all higher orders to vanish. However, if the bases of the SDFs are chosen such that $[\hat{\sigma}_\alpha, \hat{\sigma}_{\alpha'}] \neq 0$, we obtain the sought-after nonlinear interactions. In this case, the spin mediates the higher-order bosonic interactions. By choosing the correct integer setting for the detuning $m$, each term can be brought into resonance separately (i.e. have the leading contribution to the dynamics), while the other terms can be eliminated via the rotating wave approximation. As such, higher detunings lead to smaller contributions from the additional terms.

This scheme can be applied to a single motional mode to produce, for example, generalised squeezing interactions or the parity operator (i.e.,~the Hamiltonian is $\propto \hat{a}^\dagger \hat{a}$). If, instead, we apply each SDF to a separate motional mode, we realise multi-mode couplings such as beam-splitter or two-mode squeezing. 
Here, we focus on the generalised squeezing interactions as described in Eq.~\eqref{eq:generalised_sqz}. We experimentally demonstrate squeezing $(n=2)$, trisqueezing $(n=3)$ and quadsqueezing $(n=4)$. Squeezing originates from the second-order term (Eq.~\eqref{eq_supp:2-order}) in the Mangus expansion by setting $m=-1$, trisqueezing from the third-order term (Eq.~\eqref{eq_supp:3-order}) by setting $m=-2$, and quadsqueezing from the fourth-order term (Eq.~\eqref{eq_supp:4-order}) by setting $m=-3$.

The squeezing and the quadsqueezing Hamiltonians are given by 
\begin{align}
    \hat{H}_{\rm eff, even}^n &= 
     \frac{i\hbar\Omega_n}{2} \hat{\sigma}_\beta (-\hat{a}^n e^{-i\theta} + \hat{a}^{\dagger n} e^{i\theta}) \label{eq_supp:sqz_quadsqz}
\end{align}
with $n = 2$ and $n= 4$, respectively.  The trisqueezing interaction with $n = 3$ is
\begin{align}
    \hat{H}_{\rm eff, odd}^n &= 
    \frac{\hbar \Omega_n}{2} \hat{\sigma}_\beta (\hat{a}^n e^{-i\theta} + \hat{a}^{\dagger n} e^{i\theta}). \label{eq_supp:trisqz}
\end{align}
There is a $\pi/2$ phase difference in the motional phase. These expressions are slightly different to those in the main text, where we omitted the changing motional phase for simplicity in Eq.~\eqref{eq:generalised_sqz}. 

 The magnitudes of these interactions are given by
\begin{align}
    \Omega_{2, 3, 4} &= \Biggl\{  \frac{\Omega_{\alpha'} \Omega_{\alpha}}{\Delta}, \frac{\Omega_{\alpha'} \Omega_{\alpha}^2}{2 \Delta^2},\frac{\Omega_{\alpha'} \Omega_{\alpha}^3}{8 \Delta^3} \Biggl\} . \label{eq_supp:mag_generalisez sqz}
\end{align}
 The spin conditioning, which results from the nested commutators relationships, is given by
\begin{align}
    \hat{\sigma}_\beta &\propto 
    \begin{cases}
    [\hat{\sigma}_{\alpha}, \hat{\sigma}_{\alpha'}]& \text{if } n~{\rm mod}~2=0\\
    \hat{\sigma}_{\alpha'}              & \text{otherwise.}
\end{cases}
\end{align}

If the spin components of the SDFs do not commute, the Magnus expansion in Eq.~\eqref{eq_supp:magnus} has an infinite number of terms. Thus, the dynamics of the Hamiltonian in Eq.~\eqref{eq_supp:2sdfs} corresponds to that of the desired generalised squeezing Hamiltonian (Eq.~\eqref{eq_supp:sqz_quadsqz},~\eqref{eq_supp:trisqz}) up to an error $\epsilon = (\Omega_\alpha/\Delta)^{n+1}$, where we assumed that the strengths of the two SDFs are balanced $\Omega_\alpha= \Omega_{\alpha'}$ and $n$ is the order of squeezing. This coherent error is due to the higher-orders terms in the Magnus expansion and decreases faster with $\Delta$ than the desired primary interaction. As such, the error can be minimised arbitrarily by increasing $\Delta$. \\
If we accept a fixed error ${\epsilon}$, and hence $\Omega_\alpha = \sqrt[n+1]{\epsilon} \Delta$, it becomes evident that since $\Omega_\alpha$ is linear in $\eta$, the Lamb-Dicke parameter, $\Delta$ also exhibits linearity with respect to $\eta$. When examining the strength of the desired interaction, $\propto\Omega_{\alpha'}\Omega_{\alpha}^{n-1}/\Delta^{n-1}$, it follows that this quantity is linear in $\eta$ irrespective of $n$. This scaling is in contrast to driving higher-order spatial derivatives of the field, for which the strength varies as $\eta^n$ (see Sec.~\ref{sec:magnitude scaling}).

\subsection{Generating the spin-dependent forces}\label{sec: generating the spin-dependent forces}

We create the two SDFs required using a M{\o}lmer-S{\o}rensen scheme. Each SDF (as defined in Eq.~\eqref{eq_supp:1sdf}) requires a bichromatic field which is composed of two tones that are symmetrically detuned from the qubit transition $\omega_\textrm{qubit}$ by $\pm\delta$. The two tones have the optical phase $\phi_1$ and $\phi_2$, respectively at the position of the ion. They define the spin-conditioning of the SDF as
$\hat{\sigma}_\alpha = \hat{\sigma}_{\phi_s} = \hat{\sigma}_{(\phi_1 + \phi_2)/2}$ as well as the SDF motional phase $\phi_\alpha = (\phi_1 - \phi_2)/2$.
Besides the spin-dependent force term, the bichromatic field also gives rise to a second spin-flip term that drives the qubit transition off-resonantly, so Eq.~\eqref{eq_supp:1sdf} is modified to:
\begin{align}  \label{eq_supp:sdf_with_carrier}
        \hat{H}_\textrm{bi} = &\frac{\hbar \Omega_\alpha}{2}\hat{\sigma}_{\phi_s} (\hat{a}e^{-i(\Delta\, t +\phi_\alpha)} + \hat{a}^\dagger e^{i(\Delta\, t + \phi_\alpha)}) \\ \nonumber &+\hbar \Omega_c \hat{\sigma}_{\phi_s - \pi/2}\cos(\delta t), 
\end{align}

where $\Omega_\alpha =\eta\Omega_c$ and the two terms do not commute. The effect of the non-commuting spin-flip term has been extensively discussed in Refs.~\citenum{roos2008ion, sutherland2019versatile,saner2023breaking}. When setting $\delta \approx \omega_\textrm{osc}$, it leads to an effective force in the $\hat{\sigma}_\phi$ basis with the strength modulated by the $J_0$ and $J_2$ Bessel functions of the first kind, i.e. $\Omega_\alpha \rightarrow \Omega_\alpha|J_0(2\Omega_c/\delta)+ J_2(2\Omega_c/\delta)| $ in Eq.~\eqref{eq_supp:1sdf}. For these experiments, we operate in a regime of small $\Omega_c$ where $|J_0 + J_2| \approx 1$. If instead $\delta \approx \omega_\textrm{osc}/2$, we create an effective spin-dependent force in the $\hat{\sigma}_z$ basis with magnitude $\Omega_\alpha \rightarrow \Omega_\alpha|J_1(2\Omega_c/\delta)+ J_3(2\Omega_c/\delta)| $ in Eq.~\eqref{eq_supp:1sdf},  
where $J_1$ and $J_3$ are again Bessel functions of the first kind. We use both of these bichromatic field configurations to create the required interaction SDFs in the desired bases.

\section{Experimental Considerations}

\subsection{Setup}

\begin{figure}[ht!]
    \centering
    \includegraphics[width=\linewidth]{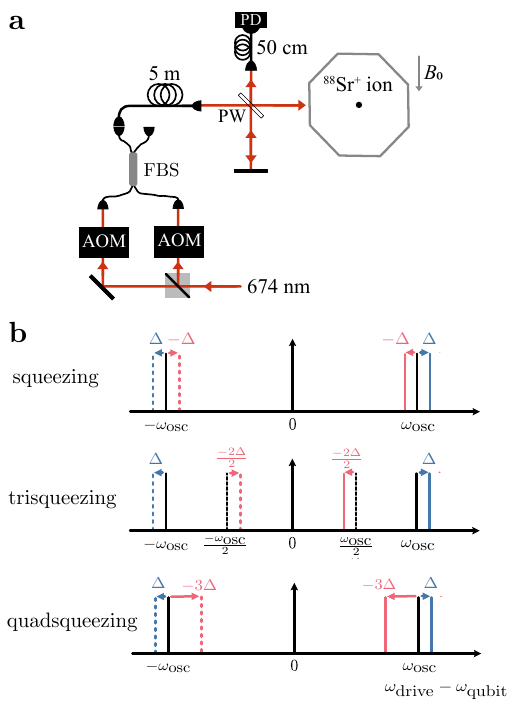}
    \caption{Overview of experimental setup. \textbf{a} Schematic of the experimental apparatus. The incoming 674\,nm beam is split into two beams. The acousto-optic modulators (AOMs) are driven with two radiofrequency tones each to generate two bichromatic fields. The two bichromatic fields are combined using a fibre beam splitter (FBS) and sent to the trap (ion not to scale). We close the resulting interferometer with a pick-off window (PW). We correct for phase fluctuations between the two beams by measuring the interference fringe intensity on a photodiode (PD) and adjusting the phase of the optical field using one of the AOMs. $B_0$ is the static magnetic field defining the quantization axis.
    \textbf{b} Frequency configuration of the bichromatic laser fields generating the nonlinear interactions. The two tones for the $\textrm{SDF}_\alpha$ are indicated in blue, while those for $\textrm{SDF}_{\alpha'}$ are in red. The two tones of each bichromatic field are symmetrically detuned from the qubit frequency one by $+\delta$ (continuous line) and one by $-\delta$ (dashed line) from the qubit frequency $\omega_{\rm qubit}$. Bichromatic field configurations for squeezing, trisqueezing and quadsqueezing are shown, where the detuning of the SDF from the motional mode changes for each interaction, $-\Delta, -2\Delta/2, -3\Delta$ respectively.
}
    \label{fig:supp_setup}
\end{figure}

In Fig.~\ref{fig:supp_setup}\textbf{a}, we show our experimental setup. For creating these interactions, we require that the spin components of the SDFs do not commute, i.e. $[\hat{\sigma}_\alpha, \hat{\sigma}_{\alpha'}] \neq 0$. As discussed in Sec.~\ref{sec: generating the spin-dependent forces}, the basis of the SDF depends on the optical phase of the tones at the position of the ion. In  Fig.~\ref{fig:supp_setup}\textbf{a}, phase fluctuations might arise from differential path length changes such as before the fibre beam splitter. We actively stabilise the optical phase, and hence the relative phase between spin components of the two SDFs, to ensure that the non-commutativity relationship is maintained throughout the experiment. The feedback loop relies on measuring the optical interference of the two beams with a photodiode (PD). We measure this interference by sampling a small fraction of light from the two beams using a pick-off mirror (PW). More details on the phase-stabilisation scheme can be found in the Supplemental Material of Ref.~\citenum{saner2023breaking}.

In Fig.~\ref{fig:supp_setup}\textbf{b}, we show the frequency configuration for implementing the two non-commuting SDFs in the case of squeezing, trisqueezing and quadsqueezing. For squeezing and quadsqueezing, the basis of the two SDFs are $\hat{\sigma}_\phi$ and $\hat{\sigma}_{\phi+\pi/2}$, respectively. To achieve this, the two tones of each of the bichromatic field are symmetrically detuned by $\delta \approx\omega_{\rm osc}$. For squeezing the exact detuning is $\delta = \omega_{\rm osc} +\Delta$ for ${ \rm SDF}_\alpha$ and $\delta = \omega_{\rm osc} -\Delta$ for ${ \rm SDF}_{\alpha'}$, while for quadsqueezing $\delta = \omega_{\rm osc} +\Delta$ and $\delta = \omega_{\rm osc} -3\Delta$, respectively. For trisqueezing, we set the bichromatic field generating ${\rm SDF}_{\alpha'}$ near resonant with $\omega_{\rm osc}/2$, with $\delta = (\omega_{\rm osc} -2\Delta)/2$ such that the spin basis is $\hat{\sigma}_z$. For ${ \rm SDF}_{\alpha}$, $\delta = \omega_{\rm osc} +\Delta$ such that its spin basis is $\hat{\sigma}_\phi$. 

\subsection{Ramp}

In the experiment, we smoothly ramp the amplitude of the interaction SDF lasers over durations $t_{\rm ramp}$ that are long compared to $1/\Delta$. The amplitude profile of the pulse $g(t)$ is given by 
\begin{equation}
g(t) = 
\begin{cases}
\sin^2(\frac{\pi t}{2t_{\rm ramp}}),\ t<t_{\rm ramp}\\
1,\ t_{\rm ramp}\leq t \leq t_f-t_{\rm ramp}\\
\sin^2(\frac{\pi (t_f -t)}{2t_{\rm ramp}}),\ t_f-t_{\rm ramp}< t<t_f.
\end{cases}
\end{equation}

\begin{figure}[ht!]
    \centering
    \includegraphics[width=\linewidth]{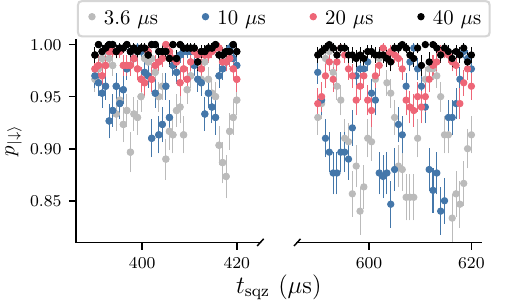}
    \caption{Example dynamics from two non-commuting SDFs (Eq.~\eqref{eq_supp:2sdfs}). While these dynamics are measured for the squeezing interaction in the $\hat{\sigma}_z$ basis, they will be similar to any other non-linear bosonic interaction in the same $\hat{\sigma}_z$ basis. The spin state is initially prepared in $\ket{\downarrow}$, before we apply the SDFs for variable durations $t_\textrm{sqz}$~\cite{fnpulseshape} and measure the probability of staying in the $\ket{\downarrow}$ state, $p_{\ket{\downarrow}}$. We repeat the measurement for different ramp durations. The amount of residual displacement, indicated by a reduction in $p_{\ket{\downarrow}}$, are suppressed as the ramp duration increases. Error bars on the data points indicate 68\% confidence intervals.}
    \label{fig:supp_loops}
\end{figure}

In the case of squeezing, we use $t_{\rm ramp} = \SI{40}{\micro \s}$ for ${\Delta/2\pi = \SI{50}{\kilo \hertz}}$ (i.e. ramp longer by a factor of 2) and for ${\Delta/2\pi = \SI{100}{\kilo \hertz}}$ (i.e. ramp longer by a factor of 4). For trisqueezing and quadsqueezing, we use $t_{\rm ramp} = \SI{80}{\micro \s}$ for $\Delta/2\pi = \SI{25}{\kilo \hertz}$ (i.e. ramp longer by a factor of 2). Due to the timescale of the decoherence channels in our system, we use a smaller value of $\Delta$ for the trisqueezing and quadsqueezing interactions to increase their strength.
The amplitude shaping of the pulse reduces its bandwidth in the frequency domain, and, hence suppresses the off-resonant driving of undesired terms in the Magnus expansion as well as the spin-flip term (see Eq.~\eqref{eq_supp:sdf_with_carrier}). Moreover, using a ramp that is long compared to $1/\Delta$ reduces the displacements in phase space as shown in Fig.~\ref{fig:supp_loops}.

We start in $\ket{\downarrow}$ and apply the squeezing interaction for variable durations $t_\textrm{sqz}$ by setting $m=-1$, $\Delta/2\pi = \SI{50}{\kilo \hertz}$ and $\SI{0.5}{\milli\watt}$, as before. The population $p_{\ket{\downarrow}}$ is measured directly after the interaction is applied. The squeezing interaction is expected to be in the $\hat{\sigma}_z$ basis, ideally leaving the initial spin state unchanged. In Fig.~\ref{fig:supp_loops}, we see that depending on the ramp duration, this is not the case. We observe periodic changes in population which indicate a residual displacement of the oscillator state in phase space. 
To obtain the squeezing interaction without any residual displacement, we need to precisely set the interaction duration where the motional state returns to the origin. However, this duration might change as a result of offsets in qubit or motional mode frequency. Using a longer ramp strongly suppresses the excursions in phase space and makes the interaction robust to any errors from the residual displacement.

\section{Numerical simulations} \label{sec:simulations}
For the simulations presented in the main text and below, we perform numerical integration of the Lindblad master equation under time-dependent Hamiltonians using the QuantumOptics.jl package in Julia~\cite{kramer2018quantumoptics}. For creating the squeezed, trisqueezed, and quadsqueezed states, we integrate the Hamiltonian in Eq.~\eqref{eq_supp:2sdfs}, without making the RWA with respect to $\omega_{\rm osc}$, and also including the spin-flip terms introduced in Eq.~\eqref{eq_supp:sdf_with_carrier}. Similarly, the probe SDF is simulated by integrating  Eq.~\eqref{eq_supp:sdf_with_carrier}, but without making the RWA with respect to $\omega_{\rm osc}$.

To recover the Fock states populations, one can drive the blue sideband corresponding to the oscillator. This interaction is equivalent to the anti-Jaynes-Cummings Hamiltonian: 
\begin{equation}
    \hat{H}_{\rm bsb} = \frac{\hbar\Omega_{\rm sb}}{2}(\hat{\sigma}_+\hat{a}^\dagger +\hat{\sigma}_-\hat{a}),
\end{equation}
which we integrate in simulation in order to analyze the results in Sec.~\ref{sec:r_value_squeezing}.

The motional decoherence \footnote{{We measured coherence times of $\SI{3}{\milli \s}$ for the Fock state superposition $(\ket{0}+\ket{1})/\sqrt{2}$.}} in our system is dominated by the heating rate $\dot{\bar{n}}_{\rm osc}=300\,(20)$~quanta/s. We introduce the heating by setting the collapse operators~\cite{harty2013highthesis} to $\sqrt{\dot{\bar{n}}_{\rm osc}}\hat{a}^\dagger$ and $\sqrt{\dot{\bar{n}}_{\rm osc}}\hat{a}$.

The Hilbert space is truncated at phonon number $50$ or $150$. The higher phonon number is especially necessary when the effect of the probe SDF is simulated.

\section{Squeezed state characterisation}\label{sec:r_value_squeezing}

In this section, we discuss how the values for the squeezing parameter $r$ were inferred for the squeezed states as shown in Fig.~\ref{fig:sqz_charact-fig2}. The ion is initialised to its $\ket{\downarrow}$ state, while its motional state corresponds to a thermal state with $\bar{n}_{\rm osc}=0.09(1)$ after cooling the motional mode close to the ground state. We then apply the squeezing interaction and the ion is left in the $\ket{\downarrow}\ket{\xi_\textrm{th}}$ state, where $\ket{\xi_\textrm{th}}$ is the squeezed (parametrised by $r$ and $\theta$, the squeezing axis) thermal state. To determine $r$, we apply a probe SDF on resonance with the oscillator frequency in the $\hat{\sigma}_x$ basis, for variable durations $t_{\rm probe}$ and power \SI{0.5}{\milli\watt}. As ${\ket{\downarrow} = (\ket{+}-\ket{-})/\sqrt{2}}$, where $\ket{\pm}$ are the eigenstates of $\hat{\sigma}_x$, the oscillator state wavefunction is split into two displaced states
\begin{equation}
    \ket{\psi} = \frac{1}{\sqrt{2}}(\ket{+}\ket{\beta, \xi_{\rm th}}-\ket{-}\ket{-\beta, \xi_{\rm th}}),
\end{equation}
where $\beta=\Omega_{\rm SDF} e^{i\phi_{\rm probe}}t_{\rm probe}/2$ quantifies the  displacement. Upon doing a projective measurement in the $\hat{\sigma}_z$ basis, by fluorescence readout, the overlap between the two displaced oscillator states, $f(\beta, \xi_{\rm th})$ is mapped on the spin. The probability of remaining in $\ket{\downarrow}$ is

\begin{equation}
    \begin{split} p_{\ket{\downarrow}}&= \frac{1 + f(\alpha,\xi_{\rm th})}{2},
    \end{split}
\label{eq_supp:wf splitting}
\end{equation}

where the overlap is defined as~\cite{lo2015spin}:
\begin{equation}
    \begin{split}
    f(\beta,\xi_{\rm th}) &=  e^{-g(\beta)h(\xi)}, \\
    g(\beta) &= |2\beta(t_\textrm{probe})|^2\left(\bar{n}_{\rm osc}+\frac{1}{2}\right), \\
    h(\xi) &= e^{2r}\cos{(\phi_{\rm probe}-\theta/2)}^2+e^{-2r}\sin{(\phi_{\rm probe}-\theta/2)}^2.
    \end{split}
\label{eq_supp:overlap splitting}
\end{equation}
The overlap depends on the relative orientation between the motional phase of the probe SDF $\phi_{\rm probe}$ and the squeezing axis $\theta$. If the two are aligned  ($\phi_{\rm probe} - \theta/2 = 0$), the splitting corresponds to the Fig.~\ref{fig:sqz_charact-fig2}\textbf{ai} inset. If instead $\phi_{\rm probe} - \theta/2 = \pi/2$, the splitting corresponds to the Fig.~\ref{fig:sqz_charact-fig2}\textbf{aiii} inset.

In the experiment, we calibrate the relative orientation ${\phi_{\rm probe} - \theta/2}$ by keeping the pulse duration for the probe SDF constant while scanning its motional phase $\phi_{\rm probe}$, see Fig.~\ref{fig:sqz_charact-fig2}\textbf{c} ``start in $\ket{\downarrow}$". We perform a fine scan over one of the peaks and fit a parabola to it in order to determine its centre. Here, the wavefunction is split about the anti-squeezed axis, which we expected to occur for $\phi_{\rm probe}$ an integer multiple of $\pi/2$. However, we observe a phase offset (see Fig.~\ref{fig:sqz_charact-fig2}\textbf{c}). We believe that the offset originates from the phase stabilisation. Importantly, the value remains constant over time and does not change with the increase in the pulse duration for the squeezing interaction and can thus be calibrated out. To split about the squeezed axis, we offset the calibrated $\phi_{\rm probe}$ by $\pi/2$. 

The model used to fit the experimental data is ${p_{\ket{\downarrow}}= (1 + Cf(\beta,\xi_{\rm th}))/2}$, where C accounts for experimental imperfections in the spin state preparation or readout. We first fit the ground-state data by setting $r=0$ in Eq.~\eqref{eq_supp:overlap splitting} and having $\Omega$ and $C$ as free parameters, which allows us to extract the strength of the probe SDF. We use this value to fit the splitting about the squeezed axis by setting $\phi_{\rm probe} - \theta/2 = 0$ and having $r$ and $C$ as free parameters.

We observe that heating during the squeezing interaction influences the splitting dynamics, resulting in an overestimate of $r$, particularly at extended squeezing durations ($t_{\rm sqz} > \SI{400}{\micro \s}$). Consequently, we attempted to incorporate the heating into our fitting analysis by assuming that the heating and the squeezing interaction are independent processes. Instead of using an initial $\bar{n}_{\rm osc}=\bar{n}_{\rm gs}=0.09(1)$, we consider that the state before applying the probe was a squeezed thermal state with ${\bar{n}_{\rm osc} = \bar{n}_{\rm gs}+\dot{\bar{n}}_{\rm osc}\, t_{\rm sqz}}$. 

\begin{figure}[ht!]
    \centering
    \includegraphics[width=\linewidth]{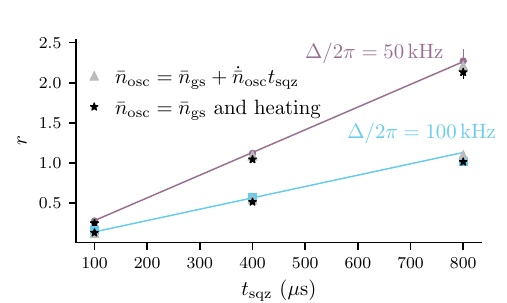}
    \caption{Verification of incorporating heating in the fitting model for extracting the squeezing parameter $r$.
    We compare the inferred $r$ values for two cases: starting in the ground state with $\bar{n}_{\rm osc} = \bar{n}_{\rm gs}$ and applying the squeezing interaction together with heating (black stars), and starting in a thermal state with $\bar{n}_{\rm osc} = \bar{n}_{\rm gs}+\dot{\bar{n}}_{\rm osc}t_{\rm sqz}$ followed by applying only the squeezing interaction (grey triangles). The second simulated case is equivalent to assuming that the heating and the squeezing interaction are independent. The splitting dynamics for the two cases are simulated and the $r$ is determined by using the same fitting procedure as for the experimental data. We overlay the simulated results to the experimental data and theory lines in Fig.~\ref{fig:sqz_charact-fig2}\textbf{b}. Error bars on the data points indicate 68\% confidence intervals based on the fit employed to estimate $r$.
    }
    \label{fig:supp_r_exp_vs_sim}
\end{figure}

We verified the validity of this assumption in simulation. We simulated the squeezing interaction followed by applying the probe SDF, see Sec.~\ref{sec:simulations}. The resulting splitting dynamics were fit in the same way as the experimental data. In Fig.~\ref{fig:supp_r_exp_vs_sim}, we compare the results for three squeezing durations, $t_{\rm sqz}$ and two magnitudes parametrised by $\Delta$. In one case, we start with a thermal state with $\bar{n}_{\rm osc} = \bar{n}_{\rm gs}+\dot{\bar{n}}_{\rm osc}t_{\rm sqz}$ and add no heating during the squeezing interaction. In the other case, we start in ${\bar{n}_{\rm osc} = \bar{n}_{\rm gs}}$ and apply heating during the squeezing interaction. As the squeezing interaction duration is increased, a slight discrepancy in the two models becomes apparent. However, this discrepancy is within the uncertainty of the experimental measurements. We plot the simulation results onto the theory and experimental data, which show good agreement, in Fig.~\ref{fig:supp_r_exp_vs_sim}.

The heating that occurs when the probe SDF is applied was not included in the fitting analysis. The splitting takes on average around $\SI{100}{\micro \s}$ and heating effects are negligible when splitting along the squeezing axis. \\

\begin{figure}[ht!]
    \centering
    \includegraphics[width=\linewidth]{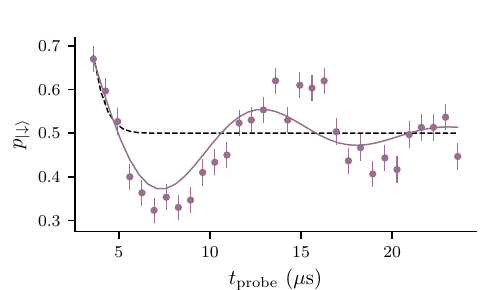}
    \caption{Splitting dynamics for applying the squeezing Hamiltonian for $t_{\rm sqz} = \SI{1000}{\micro \s}$ with $\Delta/2\pi = \SI{50}{\kilo \hertz}$ and $\SI{0.5}{\milli \watt}$ for each SDF. We show the fitting to the data using the model described above (dashed black line) and the free parameter simulation (continuous purple line). Error bars on the data points indicate 68\% confidence interval.
    }
    \label{fig:supp_1000us}
\end{figure}

In Fig.~\ref{fig:sqz_charact-fig2}\textbf{b}, we observe that the for $\Delta/2\pi = \SI{50}{\kilo \hertz}$, for $t_{\rm sqz}= \SI{1000}{\micro \s}$, the inferred $r$ of above 2 has an error bar larger than the other experimental points. Here, the fitting model does not properly describe the splitting dynamics. In Fig.~\ref{fig:supp_1000us}, we show the analysed experimental data for $t_{\rm sqz}= \SI{1000}{\micro \s}$ including the fitting and simulation. The oscillations are predicted by the simulation and they are due to the higher-order terms in the Magnus expansion becoming significant. Their effect can be mitigated by increasing $\Delta$, which would reduce the overall strength if $\Omega_{\alpha},\Omega_{\alpha'}$ are not adjusted accordingly. \\

An alternative approach to deduce the $r$ value involves extracting Fock state populations by analysing the dynamics of a sideband interaction. Subsequently, a model is used to fit the Fock state populations from the resulting dynamics and thus infer $r$, again using a model for the Fock state distribution for a given squeezed state. We did not use this method in our study due  to heating which occurs during squeezing, which complicates the resulting Fock state distribution. However, we do extract the Fock state populations and compare them to simulation.

\begin{figure}[ht!]
    \centering
    \includegraphics[width=\linewidth]{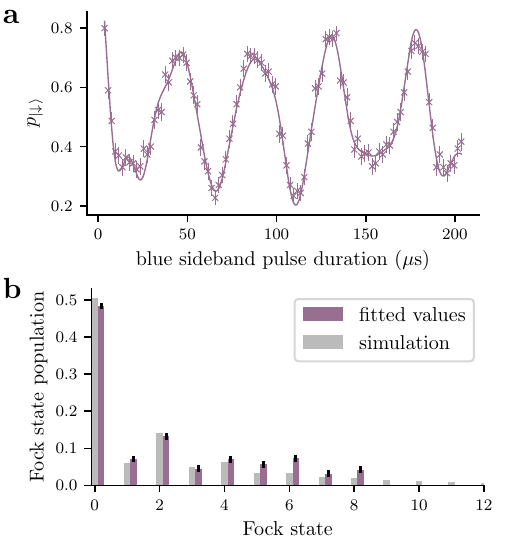}
    \caption{Fock state analysis of squeezed states.  \textbf{a}, Blue sideband dynamics. We apply a blue sideband interaction for a variable duration for a squeezed state with $r=1.09(4)$. Fock state populations are inferred by fitting an unconstrained model to dynamics. \textbf{b}, Fock state population distribution. We plot the histogram of the determined Fock state populations (purple vertical bars) and compare them to simulation (grey vertical bars). Error bars on the data points indicate 68\% confidence intervals from the shot noise for {\bf a} or based on the fit employed to estimate the Fock state populations {\bf b}.}
    \label{fig:supp_sb_analysis}
\end{figure}

We drive the motional blue sideband and fit the dynamics in Fig.~\ref{fig:supp_sb_analysis}\textbf{a} to an unconstrained model~\cite{burd2019quantum}. The purple vertical bars in Fig.~\ref{fig:supp_sb_analysis}\textbf{b} are the inferred Fock state populations. The grey vertical bars are obtained by simulating the squeezing interaction and including the heating rate the corresponding experimental parameters.

\section{Unitarity of the interactions}
An important aspect of our method that was not presented in the main text is its unitarity. Our interaction is unitary, in contrast to approaches relying on dissipative processes for generating nonlinear interactions~\cite{kienzler2015quantum}. The unitarity enables the interaction to 
be concatenated or arbitrarily placed within a single circuit, making it suitable for continuous variable quantum computing.

\begin{figure}[ht!]
    \centering
\includegraphics[width=\linewidth]{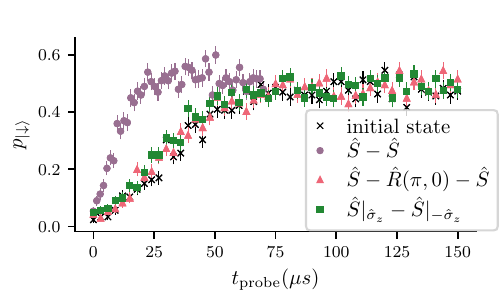}
    \caption{Verifying the unitarity of the squeezing interaction. To confirm its unitarity, we use its spin-conditioning property and the control that we have over the basis of the squeezing. We apply a probe SDF for a variable duration and measure the probability of the $\ket{\downarrow}$ state $p_{\ket{\downarrow}}$. We insert a $\pi$-pulse on the spin, $\hat{R}(\theta'=\pi,\phi'=0)$, between two squeezing interactions $\hat{S}-\hat{R}(\pi,0)-\hat{S}$ (orange triangles) or add a $\pi$ phase to the spin basis of the second squeezing pulse $\hat{S}|_{\hat{\sigma}_z}-\hat{S}|_{-\hat{\sigma}_z}$ (green squares) and confirm that the resulting splitting dynamics match that of the initial thermal state (black crosses). We also plot the data for two consecutive squeezing pulses (purple circles) which results in a squeezed state. Error bars indicate 68\% confidence intervals.}    \label{fig:supp_unitary_inverse}
\end{figure}

In this section, we experimentally investigate and verify the unitarity of our interactions (Fig.~\ref{fig:supp_unitary_inverse}). Specifically, we demonstrate that applying the spin-dependent squeezing interaction $\hat{S}$ followed by its adjoint $\hat{S}^\dagger$ to an initial state $\ket{\uparrow, \bar{n}_{\rm osc}}$ leaves the state unchanged. The sequence, implemented as two consecutive pulses, is compared across three settings against the thermal state where no squeezing pulses are applied.

In the first sequence, we apply two identical squeezing pulses for \SI{200}{\micro \second} each, i.e. $\hat{S}-\hat{S}$, resulting in a squeezed state. In the second sequence, a $\pi$-pulse on the spin, $\hat{R}(\theta'=\pi,\phi'=0)$, is introduced between the two squeezing pulses, $\hat{S}-\hat{R}(\pi,0)-\hat{S}$. Due to the spin-conditioning of the interaction, the spin-flip induced by the $\pi$-pulse transforms the second squeezing interaction into its adjoint. This effectively reverses the effect of the first squeezing interaction and restores the state to the initial thermal state. Note, we initialise the second sequence in $\ket{\downarrow, \bar{n}_{\rm osc}}$ to keep the readout consistent.

In the third sequence, the spin basis of the second squeezing pulse is changed by $\pi$ by changing the phase of one of the SDFs by $\pi$ such that the sequence is $\hat{S}|_{\hat{\sigma}_z}-\hat{S}|_{-\hat{\sigma}_z}$. Once again, this transforms the second squeezing interaction into its adjoint, resulting in the final state returning to the initial thermal state. 

Each sequence is followed by the default analysis explained in Fig.~\ref{fig:sqz_charact-fig2}\textbf{a}.

\section{Wigner function}
\begin{figure}[ht]
    \centering
    \includegraphics[width=\linewidth]{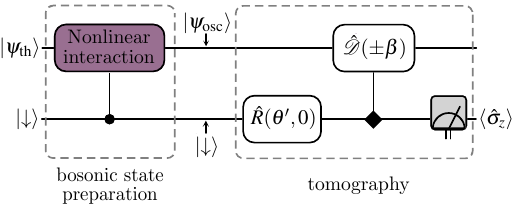}
    \caption{Circuit diagram for bosonic state preparation and tomography. The nonlinear interaction, conditioned on $\hat{\sigma}_z$ (circle), is applied to the system initialised in a thermal state and spin down $\ket{\psi_{\rm th}, \downarrow}$. The characteristic function $\chi(\beta)$ of the resulting oscillator state $\ket{\psi_{\rm osc}}$ is measured during the tomography step. The rotation $\hat{R}(\theta', \phi'=0)$ that is applied to $\ket{\downarrow}$ determines if the real or the imaginary part of $\chi(\beta)$ is measured. The displacement is conditioned on $\hat{\sigma}_x$ (diamond), which influences its orientation $(\pm\beta)$. Finally, the spin state in measured in the $\hat{\sigma}_z$ basis.}
    \label{fig:supp_pulse_charact_func}
\end{figure}
\begin{figure*}[ht]
    \centering
    \includegraphics[width=\linewidth]{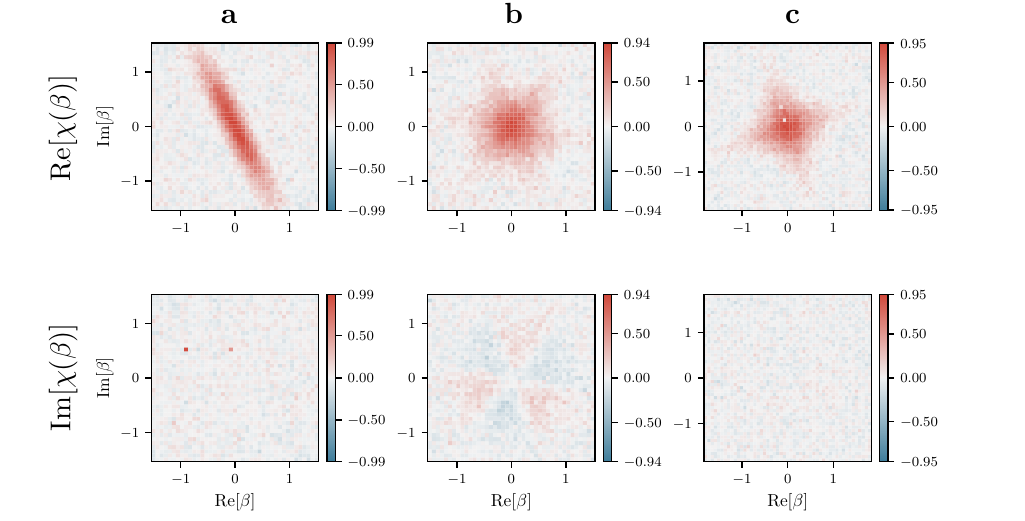}
    \caption{Measured characteristic functions for the experimentally realized generalized squeezed states. For each state, we measure the real ${\rm Re}[\chi(\beta)]$ and the imaginary part ${\rm Im}[\chi(\beta)]$. {\bf a}, For the squeezed state, the imaginary part vanishes. {\bf b}, For the trisqueezed state, the real and imaginary part have a star shaped pattern with six features. The imaginary part oscillates between positive and negative values. The resulting Wigner function has the expected triangular shape. {\bf c}, For the quadsqueezed state, the imaginary part vanishes.}
    \label{fig:supp_charact_func}
\end{figure*}
The Wigner quasiprobability function $W(\gamma)$\cite{wigner1932on} describes the wavefunction $\ket{\psi_{\rm osc}}$ of harmonic oscillator in the position-momentum phase space ($x$, $p$) expressed as the complex variable $\gamma = {\rm Re}[\gamma]+ i {\rm Im}[\gamma] \equiv x + i p$. The Wigner function fully characterises a state. It is defined as the Fourier transform of the characteristic function $\chi(\beta)= \bra{\psi_{\rm osc}}\hat{\mathcal{D}}(\beta) \ket{\psi_{\rm osc}} $ with the displacement operator $\hat{\mathcal{D}}(\beta) = e^{\beta \hat{a}^\dagger -\beta^\ast \hat{a}} $ whose argument $\beta = \beta_r + i \beta_i$ is the complex displacement variable. Thus,
\begin{align}
    W(\gamma) &= \frac{1}{\pi^2}\int  \chi(\beta) e^{\gamma \beta^\ast - \gamma^\ast \beta}d^2 \beta\\
    \longleftrightarrow W(x,p) &= \frac{2}{\pi^2}\int \int \chi(\beta_r, \beta_i) e^{2i (x \beta_i - p \beta_r)} d\beta_r d\beta_i.  
\end{align}

As demonstrated in Ref.~\citenum{fluhmann2020direct, matsos2023robust}, the real and imaginary parts of the characteristic function of a harmonic oscillator in a trapped ion system can be measured directly by applying a probe SDF which creates the displacement $\hat{\mathcal{D}}(\beta)$ (see Fig.~\ref{fig:supp_pulse_charact_func}). 

For our experiments, after the spin-dependent nonlinear interaction is applied, the system is left in the state $\ket{\psi_{\rm osc},\downarrow}$. The real part of the characteristic function ${\rm Re}[\chi(\beta)]$ is inferred by omitting the single-qubit rotation and directly applying the SDF conditioned on $\hat{\sigma}_x$, i.e. setting $\theta' =0$ for the single-qubit rotation $\hat{R}(\theta'=0, \phi'=0)$ and then, measuring in the $\hat{\sigma}_z$ basis. For the imaginary part ${\rm Im}[\chi(\beta)]$, we rotate the spin state in an eigenstate of $\hat{\sigma}_y$ with the rotation $\hat{R}(\theta'=\pi/2, \phi'=0)$ and then apply the SDF followed by a measurement in $\hat{\sigma}_z$. If the force is applied for a duration $t_{\rm probe}$, the resulting displacement is given by $\beta = \Omega_{\rm SDF} e^{i\phi_{\rm probe}}t_{\rm probe}/2$ where $\phi_{\rm probe}$ denotes the motional phase of the SDF and $\Omega_{\rm SDF}$ its strength. We vary both $t_{\rm probe}$ and $\phi_{\rm probe}$ to sample $\beta$ in the complex plane. The experimentally measured characteristic functions used to reconstruct the Wigner functions in the main text are shown in Fig.~\ref{fig:supp_charact_func}.

We measure 41 settings for each of ${\rm Re}[\chi(\beta)]$ and ${\rm Im}[\chi(\beta)]$, which are combined to get $\chi(\beta)$.
To approximate the Fourier transform of $\chi(\beta)$, we zero-pad the measured data by 200 points on every side resulting in a grid size of $441 \times 441$ and perform a discrete Fourier transform.
We do not account for the bias parameter that corresponds to state preparation and measurement errors in the spin state. We measure the entire extent of the characteristic function without making assumptions about its hermiticity, as opposed to only measuring half of the complex plane. Our reconstruction technique could also be improved by employing maximum-likelihood estimation, either directly on $\chi(\beta)$ or the Wigner function\cite{matsos2023robust}.

\section{Scaling of the nonlinear interaction strength} \label{sec:magnitude scaling}
The implementation of squeezing, trisqueezing, and quadsqueezing interactions could also be driven by higher-order terms in the Lamb-Dicke expansion~\cite{wineland1998experimental} i.e. higher-order spatial derivatives in the field. In Eq.~\eqref{eq_supp:sdf_with_carrier} we only consider the first-order term in the Lamb-Dicke expansion. However, higher-order terms are present which can drive higher order motional sidebands. Thus, by using two tones to create a bichromatic field, one can obtain the generalised squeezing interactions. For example, this method has been used to implement squeezing in Ref.~\citenum{meekhof1996generation}. The magnitudes of these interactions $\Omega_{\eta^n}$, where $n$ is, as before, the order of the interaction, are
\begin{align}
    \Omega_{\eta^2, \eta^3, \eta^4} &=  \Biggl\{  \frac{\Omega_c \eta^2}{2!}, \frac{\Omega_c \eta^3}{3!}, \frac{\Omega_c \eta^4}{4!} \Biggl\}.  \label{eq_supp:mag_n-sb}
\end{align}
We want to compare these magnitudes to those used to generate the generalised squeezed states in Fig.~\ref{fig:wigner}. We assume the same power is available for both methods, i.e., if the two non-commuting SDFs method uses \SI{0.5}{\milli \watt} for each SDF, then the higher motional sideband method, which only requires one bichromatic field, uses $2\times \SI{0.5}{\milli \watt} = \SI{1}{\milli \watt}$.

\begin{figure}[ht]
    \centering
    \includegraphics[width=\linewidth]{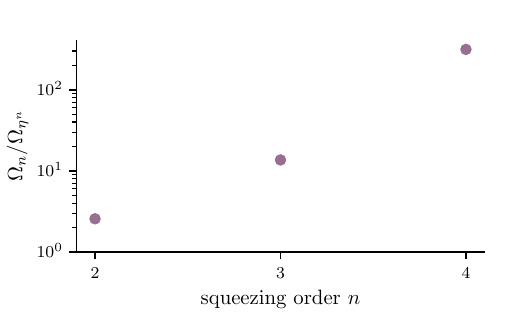}
    \caption{Comparison of the strength of generalised squeezing interactions. We compare the method demonstrated in this work, with magnitude $\Omega_n$, to the method of driving higher-order spatial derivatives in the field, magnitude $\Omega_{\eta^n}$. This comparison is done specifically for the parameters used to generate the states in Fig.~\ref{fig:wigner}. We assume the same total amount of laser power is used in both cases.
    }
    \label{fig:supp_mag_compare}
\end{figure}

In Fig.~\ref{fig:supp_mag_compare}, we plot the ratio of $\Omega_n$ and $\Omega_{\eta^n}$ (Eq.~\eqref{eq_supp:mag_generalisez sqz} and ~\eqref{eq_supp:mag_n-sb}) for the parameters (detuning $\Delta$, power) used to generate the states in Fig.~\ref{fig:wigner}. The detuning $\Delta$ can be adjusted to increase the magnitude $\Omega_n$. Hence, this plot does not give a comprehensive view of the favorable scaling for the method demonstrated in this paper.  However, it shows that, using the same power, our trisqueezing and quadsqueezing interactions are more than 10 and a 100 times stronger, respectively, than interactions using the higher-order derivatives of the field. Without this increase in strength, these interactions would have been unfeasible in our system due to the decoherence effects.

\section{Comparison of ideal and effective generalised squeezing interactions}
We want to evaluate, in simulation, how a state created using Eq.~\eqref{eq_supp:2sdfs}, with the carrier included (Eq.~\eqref{eq_supp:sdf_with_carrier}) compares to a generalized squeezed state created using the ideal interaction shown in  Eq.~\eqref{eq_supp:sqz_quadsqz}. We consider the overlap between two states given by
\begin{equation}
    F(\rho,\sigma) = \left({\rm tr}\sqrt{\sqrt{\rho}\sigma\sqrt{\rho}}\right)^2,
\end{equation}
where $\rho$ and $\sigma$ are the density matrices of the two quantum states, respectively. For simulating the states using two non-commuting SDFs, we use the independently measured parameters employed to generate the states in Fig.~\ref{fig:wigner}.
We then optimize the strength of the ideal generalised squeezing interaction to maximise the overlap between the states. We first consider the case where the oscillator is initialised to its ground state, and that there is no heating during the interaction. Here, ${1-F<\num{9e-4}}$, while for starting in $\bar{n}_{\rm osc}=0.09$ and including $\dot{\bar{n}}_{\rm osc} = 300$ quanta/s during the interactions, ${1-F<\num{1.6e-3}}$. These values apply to the squeezed, trisqueezed, and quadsqueezed states.

\label{suppl_material}
\end{document}